# Segmentation Robustness and Predictive Utility in Glioblastoma Radiomics: Evidence for a Trade-off in Survival Modelling

Mariya Miteva[1], Maria Nisheva-Pavlova[2]

*Faculty of Mathematics and Informatics – Sofia University St. Kliment Ohridski, 5 James Bourchier Blvd., Sofia 1164, Bulgaria*

## Abstract

Radiomic biomarkers derived from magnetic resonance imaging (MRI) have been widely investigated as non-invasive tools for tumor characterization and prognostic modeling in glioblastoma (GBM). However, their clinical translation remains limited, in part due to sensitivity to tumor segmentation variability. In this study, we systematically investigate the relationship between feature robustness and predictive utility in GBM survival modeling using the University of Pennsylvania Glioblastoma Imaging, Genomics, and Radiomics (UPENN-GBM) cohort. A total of 4,752 radiomic features were obtained from multiparametric MRI across three tumor subregions: enhancing tumor (ET), peritumoral edema (ED), and necrotic core (NC). Feature robustness was quantified using the intraclass correlation coefficient (ICC) based on the automatic and expert-refined segmentation versions. Among features with valid ICC estimates, 48.1% were classified as robust. Survival prediction was evaluated using cross-validation with Coxnet, Random Survival Forest, and Gradient Boosting Survival Analysis models. In this cohort, radiomic feature inclusion showed no consistent improvement over the clinical baseline, and robustness filtering produced no detectable performance gain. Model-selected features were less robust than the overall feature pool, indicating a lack of enrichment for robustness. These findings suggest that robustness alone is not a reliable criterion for feature selection in radiomics-based survival modelling.

## Introduction

Radiomics has emerged as a powerful computational framework for extracting quantitative imaging biomarkers from routinely acquired medical images [1, 2]. Since its formalization in the early 2010s, radiomics has aimed to transform medical images into high-dimensional quantitative descriptors that capture tumor phenotype beyond visual interpretation [3]. Standardization efforts such as the Image Biomarker Standardisation Initiative (IBSI) have supported the development of reproducible pipelines by providing harmonized definitions of features and guidelines for image processing and reporting [4-6]. As a result, promise across a range of clinical applications, although its translation into robust and generalizable clinical tools remains limited [7-9].

[1] Corresponding author. Tel.: +359-899-661-730
*E-mail address*: mmiteva@fmi.uni-sofia.bg
[2] Tel.: +359-879-243-434
*E-mail address*: marian@fmi.uni-sofia.bg

In neuro-oncology, radiomic analysis of MRI has been widely investigated in GBM, the most aggressive primary brain tumor in adults. According to the World Health Organization (WHO), GBM is an aggressive isocitrate dehydrogenase (IDH)-wildtype diffuse astrocytic glioma with features such as microvascular proliferation or necrosis [10]. GBM is characterized by marked intratumoral heterogeneity and highly infiltrative growth patterns that extend beyond the visible lesion on conventional MRI, complicating accurate delineation and treatment planning. Despite standard combined treatment including surgery, radiotherapy and temozolomide, patient outcomes remain poor, with median overall survival typically ranging from 12 to 15 months [11, 12]. Radiomics offers the potential to capture spatially complex imaging phenotypes associated with infiltration, tissue composition, and regional variation, and has demonstrated utility in survival prediction and molecular characterization. Survival prediction in GBM involves right-censored outcomes and therefore requires modelling frameworks that explicitly account for incomplete event-time observations.

However, the clinical translation of radiomics remains limited, largely due to concerns regarding reproducibility [13, 14]. Radiomic features are sensitive to multiple sources of variability across the imaging pipeline, including acquisition parameters, preprocessing methods, reconstruction algorithms, and tumor segmentation. Among these factors, segmentation variability is particularly critical in brain tumors, where tumor boundaries are often ambiguous and heterogeneous across imaging sequences [5, 15-17]. To address this issue, many studies incorporate reproducibility analysis using metrics such as the ICC, with features exceeding predefined thresholds (e.g., ICC $\geq$ 0.80 - 0.85) considered robust and retained for modeling [18]. Nevertheless, this approach assumes that more reproducible features are also more informative for prediction, although this is rarely evaluated in survival modelling settings with censored data [19].

In this context, understanding the relationship between feature robustness and predictive utility requires evaluation within appropriately specified survival modelling frameworks [20]. While prior studies have investigated radiomic feature robustness or survival prediction in GBM separately, the extent to which robustness translates into prognostic performance under censoring-aware evaluation remains unclear [21, 22].

In this study, we present a systematic and comprehensive analysis of this relationship in the context of GBM by quantifying feature robustness across the automatic and expert-refined segmentation versions and rigorously evaluating its impact on survival prediction within censoring-aware modelling frameworks. We further investigate the robustness characteristics of model-selected features to better understand how reproducibility relates to predictive relevance. Through this integrative approach, we aim to determine whether segmentation robustness can serve as a reliable and clinically meaningful proxy for predictive utility in survival modelling settings, providing a direct empirical evaluation of a commonly assumed but rarely tested relationship in radiomics.

# Methods

To provide an overview of the study design and analytical workflow, the main steps of the pipeline are illustrated in Fig. 1. The workflow encompasses radiomic feature extraction from multiparametric MRI using the automatic and expert-refined segmentation versions, followed by quantification of feature robustness using the ICC. Multiple feature sets were then constructed, including clinical variables alone, all radiomic features, and subsets filtered based on segmentation robustness. Survival prediction was evaluated within a censoring-aware framework using repeated cross-validation, and the relationship between feature robustness and predictive utility was systematically analysed.

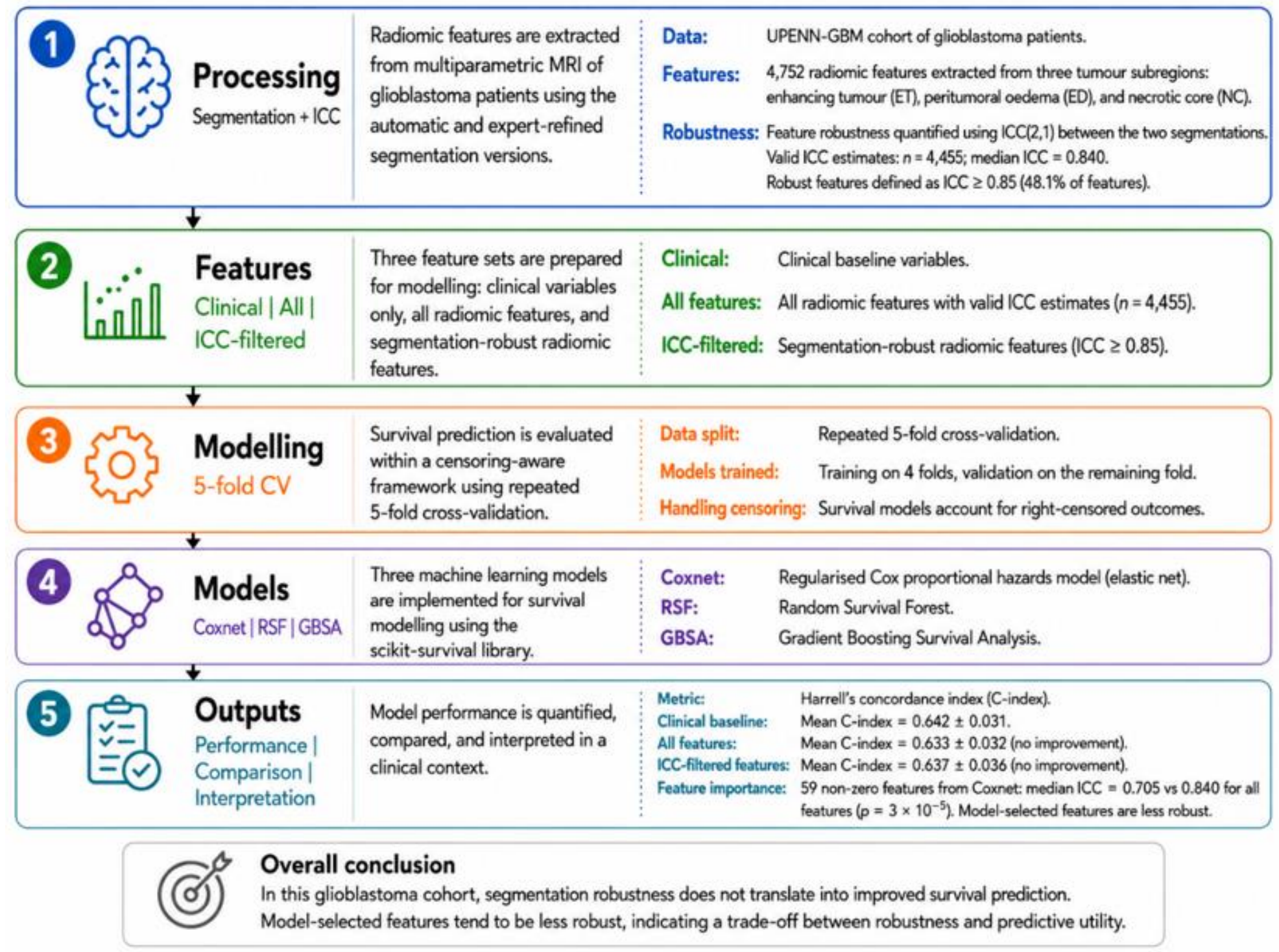

**Fig. 1.** Study design and analysis pipeline. Radiomic features were extracted using automatic and expert-refined segmentations, and robustness was quantified using ICC. Multiple feature sets were evaluated in a censoring-aware survival modelling framework with repeated cross-validation. Performance was assessed using the C-index.

### Study cohort and data assembly

This study analyzed patients with GBM from the UPENN-GBM dataset [23], publicly available through The Cancer Imaging Archive (TCIA) [24]. The dataset comprises preoperative multiparametric MRI, including T1-weighted (T1), contrast-enhanced T1-weighted (T1CE), T2-weighted (T2), and fluid-attenuated inversion recovery (FLAIR) sequences, diffusion tensor imaging (DTI), and perfusion-weighted imaging (DSC) sequences, as well as clinical annotations, molecular markers and radiomic features derived from standardized image-processing pipelines. MRI examinations were acquired prior to surgical intervention, enabling radiomic characterization of untreated tumors.

Clinical variables included age at scan, sex, IDH1 mutation status, MGMT promoter methylation status, Karnofsky performance status (KPS), extent of resection (>90%), time since baseline preoperative assessment, and pseudoprogression-related score (PsP_TP_score), subject to availability. All variables were aligned across modalities using unique subject identifiers.

The final study cohort consisted of 585 patients with matched clinical, radiomic and survival data after data alignment and preprocessing. For each patient, a total of 4,752 radiomic features were analyzed, obtained from multiparametric MRI across three tumour subregions (ET, ED and NC), as well as from diffusion- and perfusion-derived parametric maps (e.g., fractional anisotropy (FA), radial diffusivity (RD), axial diffusivity (AD), peak height (PH), percent signal recovery (PSR) and relative cerebral blood volume (ap-rCBV), resulting in a high-dimensional feature space substantially exceeding the number of samples.

Survival was modeled as a time-to-event outcome. Event indicators were derived from *Survival_Status* and cross-checked against *Survival_Censor*, with deceased patients treated as events and all other cases as censored observations. Time-to-event or follow-up time was obtained from *Survival_from_surgery_days_UPDATED* or *Time_since_baseline_preop*, as appropriate. Patients with missing survival time or event annotation were excluded from survival modeling. Survival outcomes were therefore represented as right-censored time-to-event data within a complete-case cohort.

**Radiomic features and robustness analysis**

Radiomic features were obtained from feature tables derived using the Cancer Imaging Phenomics Toolkit (CaPTk) [25] and provided with the dataset. For each segmentation pipeline, feature tables were parsed and merged into subject-level matrices, preserving feature identifiers encoding imaging modality and tumour subregion. The resulting feature space comprised 4,752 radiomic features per patient.

Features were grouped into radiomic families, including intensity-based descriptors, histogram features, shape and morphologic features, and higher-order texture features derived from gray-level matrices, including the gray-level co-occurrence matrix (GLCM), gray-level run-length matrix (GLRLM), gray-level size zone matrix (GLSZM), gray-level dependence matrix (GLDM) and neighboring gray-tone difference matrix (NGTDM).

Segmentation robustness was quantified using ICC(2,1), representing a two-way random-effects model for absolute agreement between measurements. ICC values were computed from paired feature measurements derived from the automatically generated *automaticsegm* masks and the expert-reviewed *segm* masks, which were manually refined when necessary. Both segmentations delineated the same tumour subregions. ICC was estimated once on the full cohort before cross-validation to improve stability; as no survival information was used, the resulting robustness mask was outcome-independent and applied to all folds. Features for which ICC could not be reliably estimated were excluded from the robustness analysis, and those with ICC $\geq 0.85$ were classified as segmentation-robust.

**Survival modelling and evaluation**

Survival prediction was evaluated using repeated 5-fold cross-validation with shuffled splits. All preprocessing steps were fitted exclusively within each training fold to prevent information leakage and then applied to the corresponding test fold. Numerical variables were imputed using the median and standardized by z-score normalization, whereas categorical variables were imputed using the most frequent category and one-hot encoded with unknown categories ignored at test time. Columns containing only missing values within a training fold were removed before model fitting.

Three classes of censoring-aware survival models were benchmarked. Coxnet was implemented as an elastic-net penalized Cox proportional hazards model with an L1-dominant penalty (l1_ratio = 0.9), a regularization path of 30 alpha values spaced logarithmically between $10^{-3}$ and 10[1], and fold-wise fitting after preprocessing. Random Survival Forest was implemented with 300 trees, min_samples_split = 10 and min_samples_leaf = 5. Gradient Boosting Survival Analysis was implemented with 200 boosting iterations, learning_rate = 0.05 and max_depth = 2. Hyperparameters were fixed across experimental conditions using standard settings to ensure consistent comparison between feature-set configurations. Three feature-set configurations were evaluated: (1) clinical variables only; (2) clinical variables combined with all radiomic features; and (3) clinical variables combined with robustness-filtered radiomic features (ICC $\geq$ 0.85). Predictive performance was quantified primarily using Harrell's concordance index (C-index) on held-out folds and summarized as mean $\pm$ standard deviation across resamples.

### Feature importance and statistical analysis

To investigate the relationship between segmentation robustness and model-derived importance, a descriptive Coxnet model was fitted on the full dataset following preprocessing using l1_ratio = 0.9, 50 data-derived alpha values (alpha_min_ratio = 0.01) and max_iter = 50,000; the $\alpha = 0.08$ solution retained 59 radiomic features, with sensitivity assessed across 41 non-degenerate path solutions. This model was used exclusively for feature inspection and not for estimating predictive performance. Features with non-zero coefficients were interpreted as selected by the penalized Cox model. The distribution of ICC values for these features was compared with that of the full radiomic feature set to assess whether model-selected features were preferentially associated with higher or lower robustness.

Subsequently, statistical analyses were performed to assess differences in feature robustness and model performance. The distribution of ICC values for model-selected features was compared with that of the full feature set using the two-sided Mann–Whitney U test. Differences in predictive performance across modelling configurations were evaluated using paired tests on cross-validation results. All tests were two-sided, and a significance level of $P < 0.05$ was considered statistically significant.

## Results

### Segmentation robustness of radiomic features

Segmentation robustness was quantified using ICC(2,1) by comparing radiomic features derived from two automatic and expert-refined segmentations. Among features with valid ICC estimates (n = 4,455), the median ICC was 0.840, and 48.1% of features exceeded the predefined robustness threshold (ICC $\geq$ 0.850). The ICC distribution was strongly right-skewed (as illustrated in Fig. 2), indicating that although many radiomic descriptors remained highly reproducible under segmentation perturbation, a substantial tail of low-robustness features persisted.

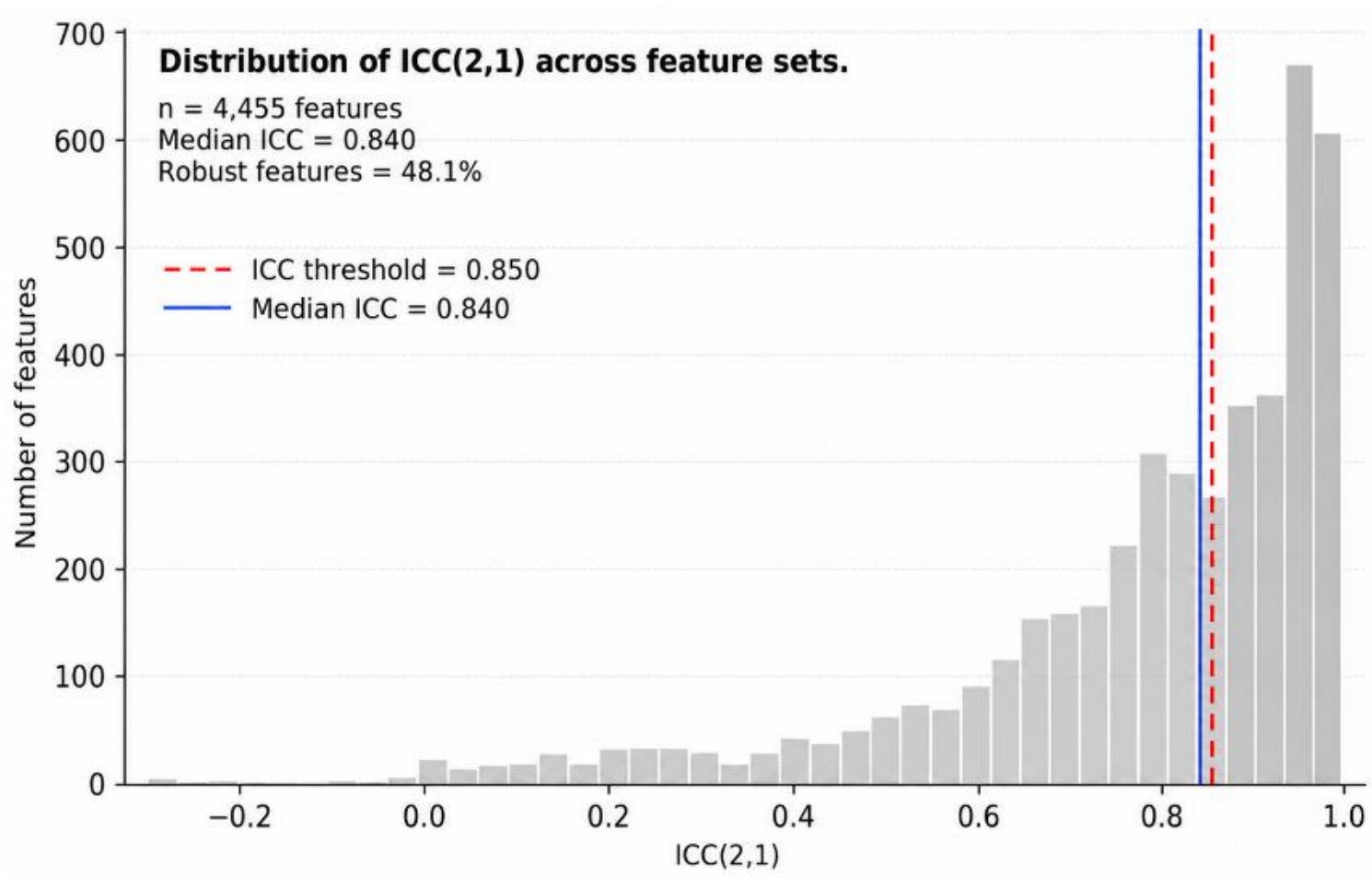

**Fig. 2.** Distribution of feature robustness across the radiomic space. Histogram of ICC(2,1) values computed for radiomic features with valid robustness estimates (n = 4,455) across automatic and expert-refined segmentations. Vertical lines indicate the prespecified robustness threshold (ICC = 0.850) and the median ICC. The distribution is right-skewed, with 48.1% of features exceeding the robustness threshold.

Feature robustness varied substantially across the radiomic feature space, indicating heterogeneous sensitivity to segmentation variability. Stratified analysis revealed systematic differences across radiomic feature families, imaging modalities and tumour compartments. Intensity-based features showed the highest robustness, followed by shape/morphological descriptors, whereas histogram and texture features showed lower agreement, suggesting greater sensitivity of higher-order spatial features to segmentation changes. Across imaging modalities, DSC-derived features were the most robust, while structural MRI sequences such as T1 and T1GD showed lower robustness, with intermediate values for DTI, FLAIR and T2. Robustness also differed across tumour compartments and derived imaging regions: perfusion-derived features showed the highest robustness rates, particularly PH (0.70), PSR (0.68) and ap-rCBV (0.66), whereas ED (0.27) and NC (0.30) were less stable; ET showed intermediate robustness (0.37). Overall, segmentation robustness depended on feature construction, imaging modality and tumour region rather than being uniformly distributed across the radiomic feature space.

## Survival prediction under censoring-aware evaluation

Survival prediction was evaluated using censoring-aware models within repeated 5-fold cross-validation with shuffled splits (5 repeats; 25 resamples in total), with performance quantified using the C-index. The clinical baseline model achieved a mean C-index of 0.642 ± 0.031. Incorporation of radiomic features did not improve predictive discrimination (0.633 ± 0.032), and restricting the feature space to segmentation-robust features (ICC ≥ 0.85) likewise did not yield a performance gain (0.637 ± 0.036). A joint robustness–predictiveness ranking strategy resulted in comparable performance (0.632 ± 0.030).

Across all modelling strategies, differences in performance between feature-set configurations were modest and largely within overlapping confidence intervals (as depicted in Fig. 3), indicating the absence of a consistent or systematic improvement associated with radiomic feature inclusion or robustness filtering. Notably, the clinical baseline remained competitive across all models and, in several instances, achieved the highest discrimination.

Performance patterns were consistent across Coxnet, Random Survival Forest and Gradient Boosting Survival Analysis, with no model demonstrating systematic superiority across feature configurations. These findings indicate that, under censoring-aware evaluation, radiomic features did not provide incremental prognostic value beyond standard clinical variables in this cohort.

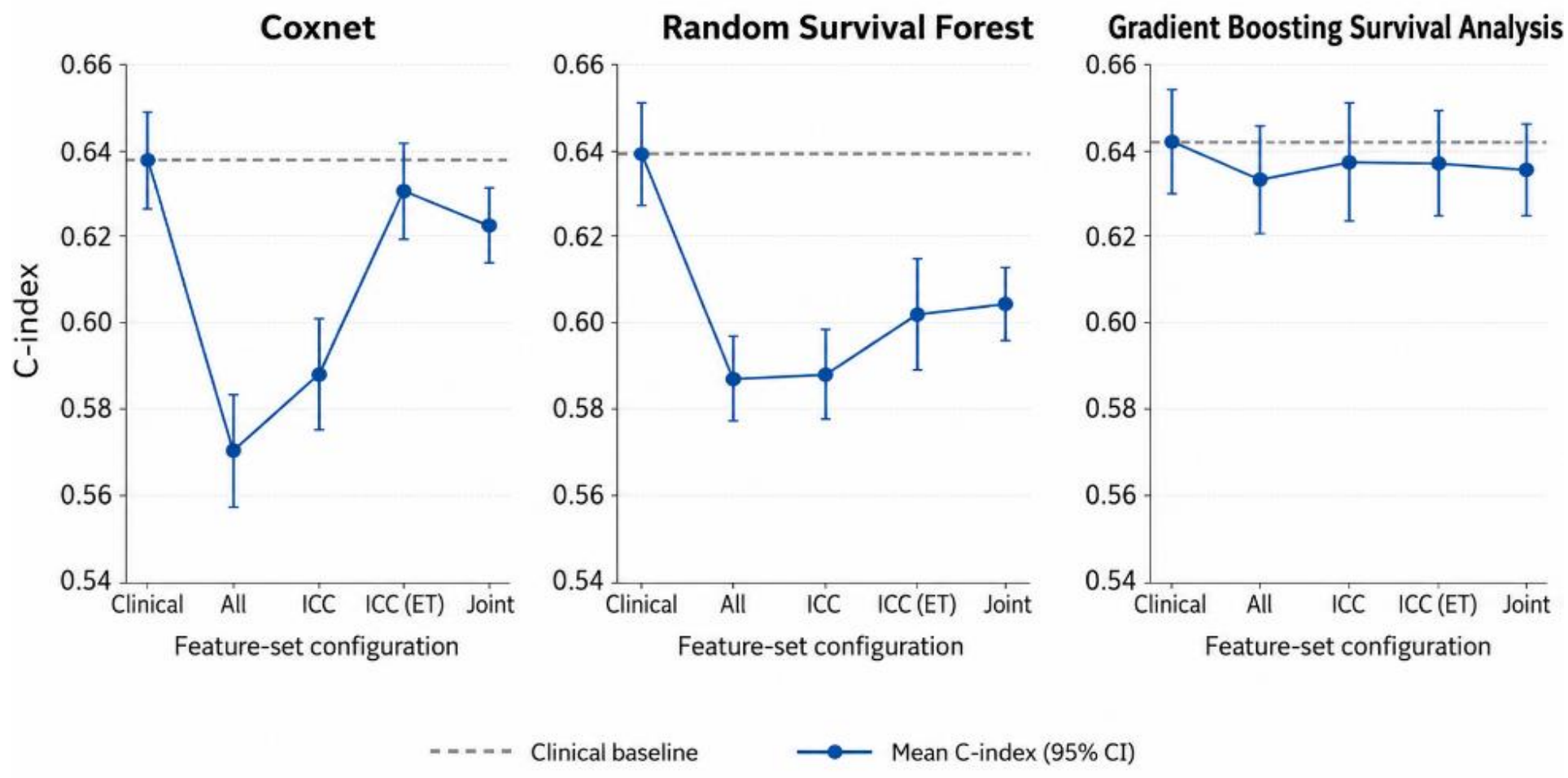

**Fig. 3**. Cross-validated prognostic performance. Mean C-index across repeated 5-fold cross-validation for Coxnet, Random Survival Forest and Gradient Boosting Survival Analysis across feature-set configurations. Error bars indicate 95% confidence intervals; dashed lines denote the clinical baseline. No consistent improvement in predictive discrimination is observed with radiomic feature inclusion or robustness filtering.

## Relationship between robustness and model-derived importance

To investigate the relationship between segmentation robustness and model-derived importance, a descriptive Coxnet model was fitted on the full dataset following preprocessing.

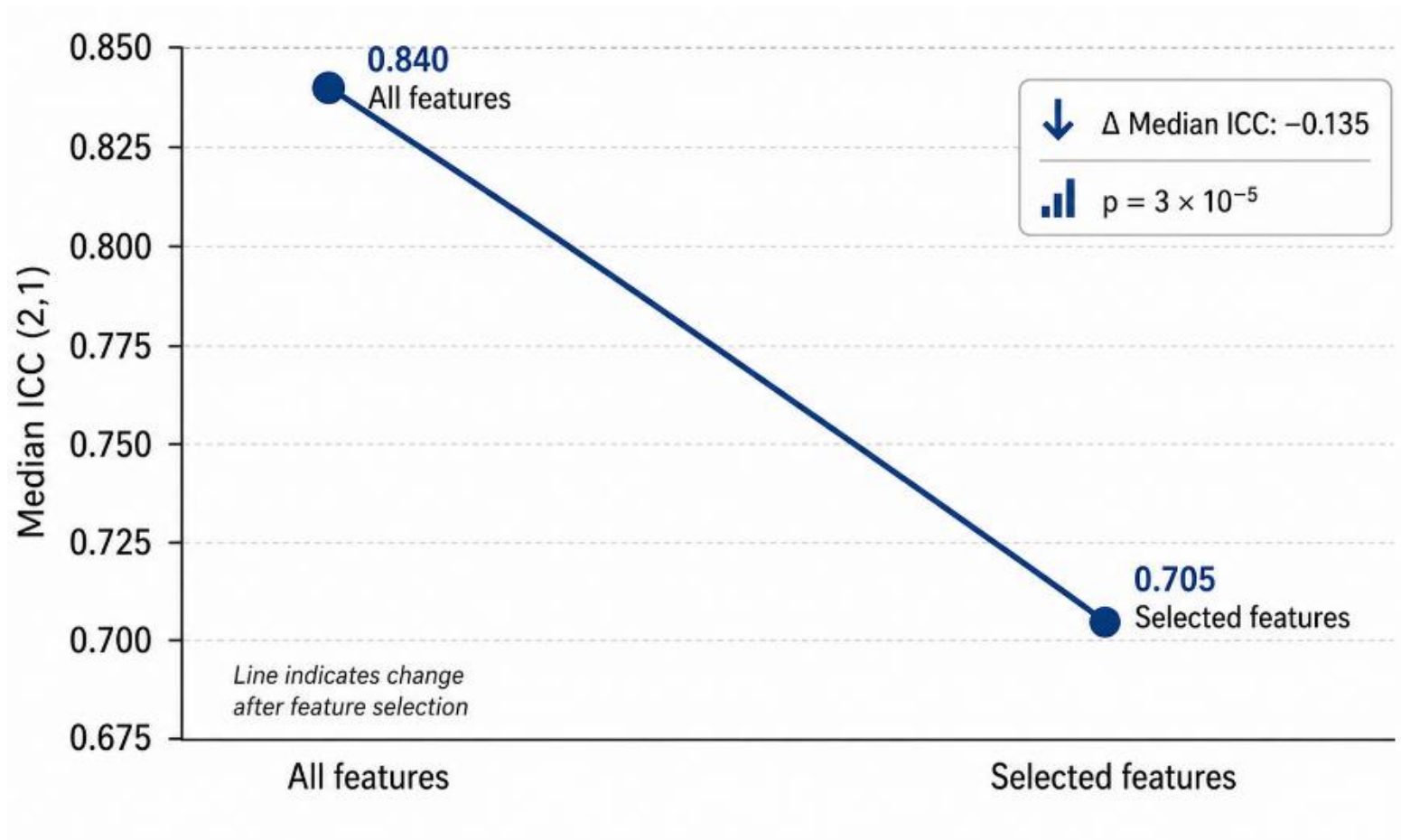

**Fig. 4.** Segmentation robustness of model-selected features. Median ICC(2,1) for all radiomic features and for features with non-zero coefficients in the Coxnet model. Model-selected features exhibit lower robustness than the overall feature pool ($p = 3 \times 10^{-5}$), indicating that predictive features are not enriched for segmentation stability.

At $\alpha = 0.08$, Coxnet retained 59 radiomic features with lower robustness than the full feature pool (median ICC 0.705 vs 0.840; $p = 3 \times 10^{-5}$; Fig. 4). A lower median ICC was also observed in 38 of 41 non-degenerate path solutions.

# Discussion

This study examined the relationship between segmentation robustness and predictive utility in radiomics-based survival modelling for GBM. Although a substantial proportion of features were classified as robust, robustness filtering did not improve prognostic performance. Models incorporating radiomic features, either unrestricted or restricted to robust features, did not outperform the clinical baseline. These findings suggest that robustness alone is not a sufficient criterion for feature selection in survival modelling [17, 26, 27].

A key observation is that features selected by predictive models were less robust than the overall radiomic feature pool, indicating a decoupling between reproducibility and predictive relevance. One possible explanation is that less robust features may capture fine-grained spatial heterogeneity or boundary-related variation that is clinically informative but sensitive to tumour delineation. In contrast, highly robust features may reflect more global intensity or structural properties that are less informative for modelling complex processes such as tumour progression and survival [8, 28].

These findings have important implications for radiomics workflows that rely on robustness filtering as a preprocessing step. Strict threshold-based filtering may remove informative features and does not necessarily improve predictive performance. Robustness should therefore be considered complementary to predictive utility rather than a surrogate for it.

This study is limited by its use of a single retrospective cohort, the absence of external validation, and fixed hyperparameters, which may restrict generalizability and model-specific performance. Future work should assess these findings in independent multicentre cohorts using external validation and nested hyperparameter tuning, while jointly considering robustness and predictive utility [5, 6, 20].

# Conclusion

In this study, we systematically investigated the relationship between radiomic feature robustness and predictive utility in survival modelling for GBM. Using automatic and expert-refined segmentations and censoring-aware modelling approaches, we found no consistent evidence that increased segmentation robustness improved prognostic performance in this cohort. The inclusion of radiomic features, whether considered in full or restricted to robustness-filtered subsets, did not consistently outperform clinical baseline models.

Notably, features selected by predictive models were, on average, less robust than the overall feature set, suggesting that prognostic signal may lie in features sensitive to segmentation variability. This challenges the common assumption that robustness is a reliable proxy for predictive relevance and instead points to a trade-off between reproducibility and utility. Overall, our findings highlight the importance of modelling strategies that consider both aspects jointly, rather than relying solely on robustness-based feature filtering, and provide evidence of a potential trade-off between reproducibility and predictive utility within the present cohort and modelling framework.

## Acknowledgements

This work was supported by the project UNITe BG16RFPR002-1.014-0004, funded by PRIDST.